# Light propagation through random hyperbolic media

Igor I. Smolyaninov [1)*], Alexander V. Kildishev [2)]

[1)] Department of Electrical and Computer Engineering, University of Maryland, College Park, MD 20742, USA
*Corresponding author: smoly@umd.edu

[2)] Birck Nanotechnology Centre, School of Electrical and Computer Engineering, Purdue University, West Lafayette, IN 47907, USA

We analyze electromagnetic field propagation through a random medium which consists of hyperbolic metamaterial domains separated by regions of normal "elliptic" space. This situation may occur in a problem as common as 9 µm light propagation through a pile of sand, or as exotic as electromagnetic field behavior in the early universe immediately after the electro-weak phase transition. We demonstrate that spatial field distributions in random hyperbolic and random "elliptic" media look strikingly different. Optical field is strongly enhanced at the boundaries of hyperbolic domains. This effect may potentially be used to evaluate the magnitude of magnetic fields which existed in the early universe.

OCIS Codes: 160.3918

Hyperbolic or "indefinite" metamaterials [1-3] are uniaxial materials with highly non-trivial electromagnetic properties. Since diagonal components of the dielectric tensor of these materials have different signs, electromagnetic field propagation inside hyperbolic metamaterials is quite unusual. For example, it was demonstrated theoretically [1-3] that there is no usual diffraction limit in a hyperbolic metamaterial. This prediction has been confirmed experimentally in refs. [4,5]. While much attention so far has been devoted to properties of "ordered" hyperbolic metamaterials, in which the optical axis direction does not change in space, it is also interesting to consider the properties of "random" hyperbolic metamaterials, which consist of multiple "hyperbolic domains" having different orientations of their optical axis. These domains may be separated by regions of "normal" (elliptic) dielectric material. Such a situation may occur in a problem as common as 9 µm light propagation through a pile of sand. Many anisotropic crystalline materials, such as silicon oxide, exhibit narrow Reststrahlen bands in the long wavelength infrared range [6]. Inside the Reststrahlen band the dielectric tensor components of these materials are negative. Since different components of the dielectric tensor pass through zero at slightly different frequencies, narrow hyperbolic bands appear at the boundaries of the Reststrahlen band. In silicon oxide this "natural" hyperbolic behaviour occurs around 9 µm. Therefore, a pile of sand at this wavelength behaves as a collection of randomly oriented hyperbolic domains separated by air gaps.

On the other hand, somewhat similar situation probably occurred in the early universe immediately after the electro-weak phase transition. According to most estimates, magnetic fields in the early universe were extremely strong. Magnetic fields up to $10^{19}$ T probably existed at the time of electro-weak phase transition [7]. This range of magnetic fields appears to be three orders of magnitude stronger than $B_c \sim 10^{16}$ T fields required to induce a superconducting vacuum state, which has been recently proposed by Chernodub [8,9]. Therefore, it is not unreasonable to suggest that superconducting vacuum domains separated by regions of "normal" vacuum did exist in the early universe. If the superconducting vacuum domains did exist at some point, they had to leave very characteristic traces. This conclusion is based on the recently demonstrated "hyperbolic metamaterial" behavior of superconducting vacuum domains [10,11]. Diffractionless field propagation through hyperbolic domains was bound to leave verifiable traces in the large scale structure of present-day universe.

Motivated by these problems, we have performed numerical simulations of field propagation through a model medium which consists of multiple randomly oriented hyperbolic domains separated by regions of normal "elliptic" space. It appears that spatial field distributions in such media look strikingly different from the field distributions in usual "elliptic" random media. According to our simulations, electromagnetic energy accumulates at the hyperbolic domain boundaries. Such energy accumulation on the domain walls at the time of electroweak phase transition probably contributed to seeding the large scale structure of present-day universe [12].

Both hyperbolic media considered here may be described as non-magnetic ($\mu$=1) uniaxial anisotropic materials having dielectric permittivity $\varepsilon_1$ in the direction perpendicular to the optical axis, and $\varepsilon_2$ along the axis. The wave equation in such a material can be written as

$$-\frac{\partial^2 \vec{E}}{c^2 \partial t^2} = \vec{\varepsilon}^{-1} \vec{\nabla} \times \vec{\nabla} \times \vec{E} \qquad (1)$$

where $\vec{\varepsilon}^{-1}$ is the inverse dielectric permittivity tensor calculated at the center frequency of the signal bandwidth. Any electromagnetic field propagating in this uniaxial material can be expressed as a sum of the "ordinary" ($E$ perpendicular to the optical axis) and "extraordinary" (vector $E$ parallel to the plane defined by the k–vector of the wave and the optical axis) contributions. We will be interested in the behavior of the extraordinary portion of the field $\varphi = E_z$ (so that the ordinary portion of the electromagnetic field does not contribute to $\varphi$). Equation (1) then yields:

$$\frac{\partial^2 \varphi}{c^2 \partial t^2} = \frac{\partial^2 \varphi}{\varepsilon_1 \partial z^2} + \frac{1}{\varepsilon_2}\left(\frac{\partial^2 \varphi}{\partial x^2} + \frac{\partial^2 \varphi}{\partial y^2}\right) \qquad (2)$$

where the optical axis is assumed to be oriented along z direction (the direction of magnetic field). While in ordinary crystalline anisotropic media both $\varepsilon_1$ and $\varepsilon_2$ are positive, in the hyperbolic metamaterials $\varepsilon_1$ and $\varepsilon_2$ have opposite signs. For example, as demonstrated in ref.[10], the unusual "hyperbolic" phase of vacuum induced by high magnetic field $B>B_c$ may be described by $\varepsilon_1 \sim 1$ and $\varepsilon_2 < 0$.

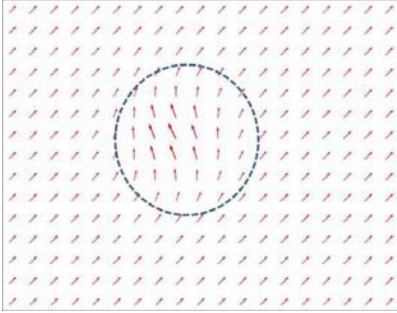

Fig. 1. (Color online) Spatial distribution of $\varepsilon_x$ and $\varepsilon_y$ components of the anisotropic dielectric tensor inside a single hyperbolic domain. In the vector representation used in this plot the tensor components are shown as a $(\varepsilon_x, \varepsilon_y)$ vector. At the domain boundary indicated by a dashed line both $\varepsilon_x$ and $\varepsilon_y$ change continuously to the vacuum $\varepsilon=1$ value.

Our numerical simulations have been performed using COMSOL Multiphysics 4.2a solver. In order to simplify calculations we have considered a two-dimensional distribution of hyperbolic domains separated by "normal" ($B<B_c$) vacuum regions. Optical axis of each hyperbolic domain had random orientation in the xy plane. An example of such a domain having optical axis oriented along x direction is shown in Fig.1. This domain has negative $\varepsilon_x$ and positive $\varepsilon_y$ components of the anisotropic dielectric tensor. At the domain boundary indicated by a dashed line in Fig.1 both $\varepsilon_x$ and $\varepsilon_y$ change continuously to the vacuum $\varepsilon=1$ value. In the "hyperbolic vacuum domain" scenario this transition occurs at the critical value $B_c$ of magnetic field. Since numerical simulations of hyperbolic media could have mesh issues, we compared our finite element modeling with analytical results, which are free of any mesh dependent issues. The analytical approach is based on the Clemmow transformations, which are mapping the free space fields onto anisotropic material space. The results from both simulations agree very well, which will be reported elsewhere. A typical mesh size which we have used was about 1/30 of the vacuum wavelength and has been verified for 1/60 ratio to ensure similar fields qualitatively and quantitatively.

First, we have studied field distribution around a single circular hyperbolic domain when a dipole source is placed in different locations inside the domain (Fig2(a,b)), and just outside the domain boundary (Fig.2(c)). The dipole radiation pattern is highly directional inside the hyperbolic domain, and in all cases we observe considerable field enhancement at the domain boundary. As a next step, we have studied field distributions in various multiple domain configurations having arbitrary elliptic shapes and random optical axis orientation.

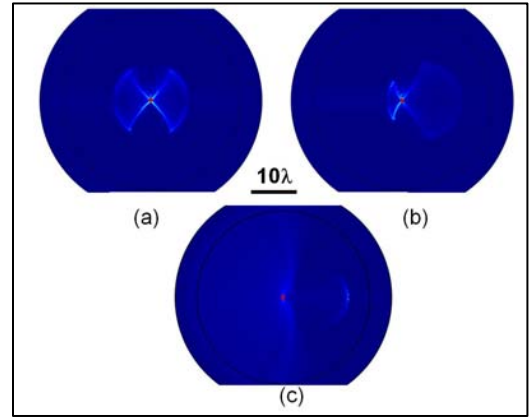

Fig. 2. (Color online) Field distributions around a single circular hyperbolic domain when a dipole source is placed in different locations inside and just outside the domain. The plots show absolute value of electric field. The dipole position is marked by a red dot.

Such numerical experiments are shown in Figs. 3 and 4(a-d). In all cases, electromagnetic field exhibits strong concentration at the domain boundaries. This effect does not depend on the domain orientation, domain shapes, and the dipole source position with respect to a given domain configuration. For example, Fig.4(a,b) shows electric field distribution in a three-domain configuration depending on the source position. Regardless of the source position, field is concentrated at the domain boundaries.

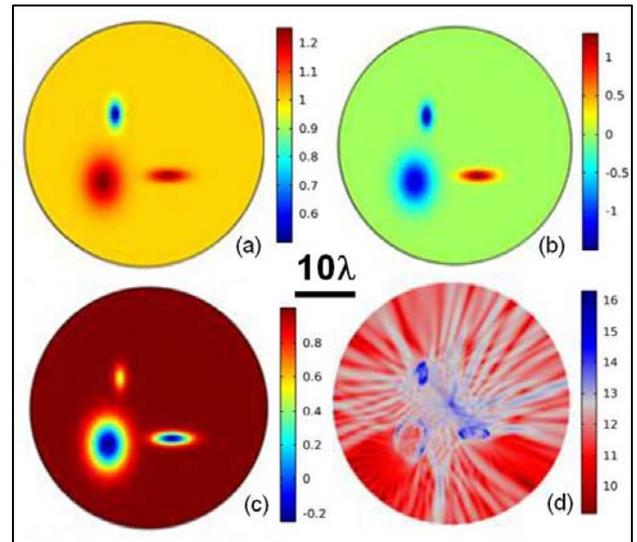

Fig. 3. (Color online) An example of field distribution in a random three-domain configuration. Panels (a-c) show $\varepsilon_{xx}$, $\varepsilon_{xy}$ and $\varepsilon_{yy}$ components of the dielectric tensor. Field distribution in logarithmic scale is plotted in panel (d). A dipole radiation source is positioned at the centre of the field of view.

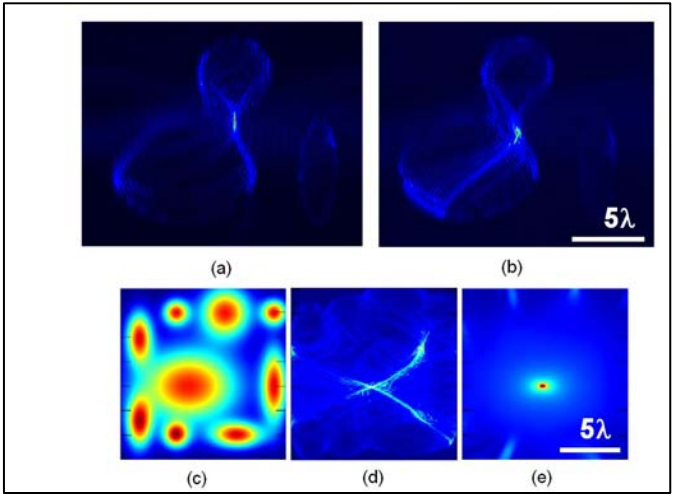

Fig. 4. (Color online) (a,b) Magnitude of electric field in a three-domain configuration depending on the source position: in all cases field is concentrated at domain boundaries. The common scale bar is given in panel (b). (c,d,e) Comparison of electric field behaviour in two identical multiple-domain configurations of hyperbolic (d) and "elliptic" (e) domains. Domain configuration in the field of view is shown in (c). The common scale bar is given in panel (e)

This result can be justified by the following analytical consideration. It appears that spatial variations of $\varepsilon_2$ may lead to a resemblance of electromagnetic "black hole" in a random hyperbolic metamaterial. Indeed, let us consider the Rindler metric

$$ds^2 = -g^2 z^2 dt^2 + dx^2 + dy^2 + dz^2 \quad (3)$$

which has a horizon at z=0. The Klein-Gordon equation in Rindler coordinates is

$$-\frac{1}{g^2 z^2}\frac{\partial^2 \varphi}{\partial t^2} + \frac{\partial^2 \varphi}{\partial x^2} + \frac{\partial^2 \varphi}{\partial y^2} + \frac{\partial^2 \varphi}{\partial z^2} + \frac{1}{z}\frac{\partial \varphi}{\partial z} = \frac{m^2 c^2}{\hbar^2}\varphi \quad (4)$$

By changing variables to $\psi = z^{1/2}\varphi$ and $\tau = gzt$, the latter equation may be rewritten as

$$-\frac{\partial^2 \psi}{\partial \tau^2} = \left(-\frac{\partial^2}{\partial x^2} - \frac{\partial^2}{\partial y^2} - \frac{\partial^2}{\partial z^2} - \frac{1}{4z^2} + \frac{m^2 c^2}{\hbar^2}\right)\psi \quad (5)$$

Eq.(5) may be emulated if the following dispersion relation is reproduced approximately inside the metamaterial:

$$\frac{\omega^2}{c^2} = k_\perp^2 + k_z^2 - \frac{1}{4z^2} + \frac{m^2 c^2}{\hbar^2} \quad (6)$$

This may be done if we manage to produce $k_z \sim z^{-1}$ behavior in the limit z→0. On the other hand, the dispersion relation of extraordinary photons inside the hyperbolic metamaterial given by eq.(2) is

$$\frac{\omega^2}{c^2} = \frac{k_\perp^2}{\varepsilon_2} + \frac{k_z^2}{\varepsilon_1} \quad (7)$$

Therefore, a spatial distribution of $\varepsilon_2$ which may be approximated as $\varepsilon_2 \sim -z^2$ would produce the desired behavior. These conditions may be realized at the hyperbolic domain boundaries. We should also point out that a more detailed study of electromagnetic field behavior at an individual interface between an elliptic and indefinite medium can be also found in ref.[13].

Finally, we have compared electromagnetic field behaviour in various sets of identical hyperbolic and "elliptic" multiple-domain configurations, as shown in Fig.4(c-e). It appears that these spatial field distributions look strikingly different. While "elliptic" domain configurations only exhibit weak lensing effect (Fig.4e), random distributions of hyperbolic domains always demonstrate strong electromagnetic energy accumulation at the domain boundaries (Fig.4d). This telltale effect could potentially be used to evaluate the magnitude of magnetic fields which existed in the early universe. The nine-domain configuration in Figs.4(c-e) has been selected randomly. We also investigated many other random configurations of up to 20 domains (the domain number being limited by mesh issues). The main result of our study – energy accumulation at domain boundaries – remains the same in all these numerical experiments.

In conclusion, we have analyzed electromagnetic field propagation through a random medium which consists of hyperbolic metamaterial domains separated by regions of normal "elliptic" space. This situation occurs when 9 μm light propagates through a pile of sand. It also probably occurred in the early universe immediately after the electro-weak phase transition. We demonstrate that spatial field distributions in random hyperbolic and random "elliptic" media look strikingly different. This effect may potentially be used to evaluate the magnitude of magnetic fields which existed in the early universe. It can also be used in such applications as studies of critical opalescence in hyperbolic metamaterials [14].